\documentclass[conference,letterpaper]{IEEEtran}

\usepackage[utf8]{inputenc} 
\usepackage[T1]{fontenc}
\usepackage{url}
\usepackage{ifthen}
\usepackage{cite}
\usepackage[cmex10]{amsmath} 

\usepackage{lmodern}
\usepackage{color}
\usepackage[pdftex]{graphicx}
\usepackage{newtxtext}
\usepackage{booktabs}
\usepackage{multirow}
\usepackage{bm}
\usepackage{bbm}
\usepackage{amsthm,amssymb,cases}
\usepackage{subfiles}
\usepackage{algorithmic}
\usepackage{algorithm}
\usepackage{mathtools}
\usepackage{mathrsfs}
\usepackage{diagbox}
\usepackage{subcaption}
\graphicspath{{./fig/}}

\newtheorem{exam}{Example}

\begin{document}
\title{Shift-Interleave Coding for DNA-Based Storage: Correction of IDS Errors and Sequence Losses}


\author{%
  \IEEEauthorblockN{Ryo Shibata}
  \IEEEauthorblockA{Graduate School of Science and Technology\\
                    Shinshu University\\
                    Nagano, Japan\\
                    Email: r-shibata@ieice.org}
  \and
  \IEEEauthorblockN{Haruhiko Kaneko}
  \IEEEauthorblockA{School of Computing\\
                    Tokyo Institute of Technology\\
                    Tokyo, Japan\\
                    Email: hkaneko@ieee.org}
}


\maketitle


\begin{abstract}
We propose a novel coding scheme for DNA-based storage systems, called the \textit{shift-interleave (SI)} coding, designed to correct insertion, deletion, and substitution (IDS) errors, as well as sequence losses.
The SI coding scheme employs multiple codewords from two binary low-density parity-check codes.
These codewords are processed to form DNA base sequences through shifting, bit-to-base mapping, and interleaving.
At the receiver side, an efficient non-iterative detection and decoding scheme is employed to sequentially estimate codewords.
The numerical results demonstrate the excellent performance of the SI coding scheme in correcting both IDS errors and sequence losses.
\end{abstract}
\section{Introduction}
Due to the growing demand for data storage, DNA-based data storage systems have gained considerable interest because of their potential for exceptionally high storage capacities and longevity\cite{dna1,dna2}.
In these systems, binary data is encoded into quaternary sequences of the four bases (i.e., A, C, G, and T).
They are then synthesized into DNA sequences for storage purposes and can later be read using sequencing technologies.
However, the processes of synthesis, storage, and sequencing can cause a DNA sequence to be missing (also known as dropout) or corrupted by insertion, deletion, and substitution (IDS) errors \cite{dna_error1,dna_error2}.
Therefore, various error correction schemes for DNA storage have been extensively studied in recent years.

In the previous works, sequence level errors (SLEs), such as sequence losses, and base level errors (BLEs), such as IDS errors, were addressed independently.
For SLEs, typical erasure correcting codes, such as the fountain codem\cite{dna_fountain}, Reed-Solomon (RS) code\cite{dna_rs}, and low-density parity-check (LDPC) code\cite{dna_ldpcmk1}, have been employed.
In contrast, for BLEs, a variety of error correcting codes have been proposed to protect each DNA sequence.
These include the RS code\cite{dna_fountain}, HEDGES convolutional code\cite{HEDGES}, modified Levenshtein code\cite{dna_lev}, LDPC code\cite{dna_ldpc1,dna_ldpc2}, and concatenated codes with outer LDPC and inner synchronization codes (such as watermarks or markers)\cite{dna_ldpcmk1,dna_ldpcmk2,dna_ldpcwm}.

Due to technological constraints in DNA storage, the length of each DNA sequence is currently limited to a few hundred bases.
This limitation forces the use of short block codes for correcting BLEs within a DNA sequence, which may not yield the desired performance.
Moreover, simultaneously addressing both SLEs and BLEs provides room for improvement in terms of enhanced performance and reduced computational complexity.
Therefore, it is important to develop a coding scheme that leverages longer codewords and efficiently manages both SLEs and BLEs with a single encoder-decoder pair.

In this study, we propose a novel coding scheme for DNA storage, called the \emph{shift-interleave (SI)} coding.
The main idea of this scheme is inspired by our previous work on binary ID-noisy channels\cite{my_tcom23,my_glo22}.
The SI coding approach involves transmitting multiple codewords using two binary LDPC codes.
At the transmitter, we utilize a three-stage encoding process---shifting, mapping, and interleaving---to transform the codewords into numerous DNA base sequences.
At the receiver, we apply an efficient, non-iterative detection and decoding method to sequentially reconstruct the original codewords, with two LDPC decoders operating in tandem.
In the proposed scheme, a synchronization structure similar to markers is incorporated into a DNA base sequence, while still achieving a suppressable rate loss, which aids in the estimation of ID errors.
Additionally, since our scheme is not constrained by codeword length, it enables the correction of erasures and substitutions using long LDPC codes.
Numerical results demonstrate the superior performance of the proposed scheme in channels affected by IDS errors and sequence losses.

\subsection{Organization}
The remainder of this paper is organized as follows.
Section \ref{sec:channel} introduces the cascaded channel model for DNA-based storage.
Section \ref{sec:proposed} details the proposed SI coding scheme.
Section \ref{sec:sim} demonstrates the finite-length performance of this scheme.
Finally, Section \ref{sec:conc} presents the conclusions.
\subsection{Notations}
Throughout the paper, we use the following notations.
For given integers $i$ and $j$ satisfying $i \leq j$, the notation $[i:j]$ denotes the set $\{i,i+1,\ldots, j\}$.
Let $\Sigma_{2}$ and $\Sigma_{4}$ denote the binary alphabet $\{0,1\}$ and DNA base alphabet $\{A,C,T,G\}$, respectively.
For a matrix $\bm{A}$, the $(i,j)$-th element is represented as $a(i,j)$,
and the $j$-th column vector is denoted by $\bm{a}_{j}$, respectively.
Moreover, using equally sized block matrices $\bm{A}^{(i,j)}, i\in [1:p], j\in [1:q]$, the matrix $\bm{A}$ can be represented as follows:
\begin{align}
  \bm{A} =
  \begin{bmatrix}
    \bm{A}^{(1,1)} & \bm{A}^{(1,2)} & \cdots & \bm{A}^{(1,q)}\\
    \bm{A}^{(2,1)} &  \bm{A}^{(2,2)} & \cdots       & \bm{A}^{(2,q)}\\
    \vdots & \vdots         & \ddots & \vdots \\
    \bm{A}^{(p,1)} & \bm{A}^{(p,2)} & \cdots & \bm{A}^{(p,q)}
  \end{bmatrix}.
\end{align}
The block column matrix $(\bm{A}^{(1,j)},\bm{A}^{(2,j)},\ldots,\bm{A}^{(k,j)})^{\top}$ is denoted by $\bm{A}^{(j)}$.
\section{DNA Channel Model}\label{sec:channel}
\begin{figure}[t]
  \centering
  \includegraphics[width=.73\linewidth]{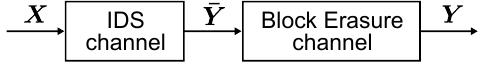}
  \caption{Cascaded IDS and block erasure channel model.}
  \label{fig:channel}
\end{figure}

We consider a simplified model of DNA-based data storage systems, where each DNA sequence can experience IDS errors and/or sequence loss.
Figure \ref{fig:channel} illustrates the channel model.
We consider the concatenation of an IDS channel with a block erasure channel.
Let $\bm{X} \in \Sigma_{4}^{n \times r}$ denote the channel input matrix.
Each column vector $\bm{x}_{j}=(x(1,j),x(2,j),\ldots,x(n,j))^{\top}$ corresponds to a DNA sequence.
In the first channel, each sequence $\bm{x}_{j}$ is independently sent through an IDS channel, where each symbol $x(i,j)$ is successively transmitted.
Each symbol $x(i,j)$ can undergo one of the following events:
\begin{enumerate}
    \item With probability $p_{\rm i}$, $x(i,j)$ is transmitted twice.
    In this case, each symbol independently experiences a substitution error with probability $p_{\rm s}$.
    \item With probability $p_{\rm d}$, $x(i,j)$ is not transmitted.
    \item With probability $1 - p_{\rm i} - p_{\rm d}$, $x(i,j)$ is transmitted.
    In this case, the symbol experiences a substitution error with probability $p_{\rm s}$.
\end{enumerate}
This process is repeated from $i=1$ to $n$, and the concatenation of the outputs forms the $j$-th column of the matrix $\bm{\bar{Y}}$, denoted as $\bm{\bar{y}}_{j} \in \Sigma_{4}^{n_{j}}$.
Note that as a substitution error model, this study employs a quaternary symmetric channel.
Then, in the second channel, each sequence $\bm{\bar{y}}_{j}$ is erased with probability $p_{\rm e}$, resulting in $\bm{y}_{j} = ?$, and $\bm{y}_{j} = \bm{\bar{y}}_{j}$ with probability $1-p_{\rm e}$,
where $\bm{y}_{j}$ denotes the $j$-th column of the output matrix $\bm{Y}$.

It should be noted that our channel model can be extended to incorporate an asymmetric error\cite{dna_ldpc1,dna_ldpc2}, shuffling-sampling\cite{dna_np}, and multi-draw models\cite{dna_ldpcwm,dna_np}.
\section{SI Coding Scheme}\label{sec:proposed}
\subsection{Overview}
Figure \ref{fig:system} shows a block diagram of the proposed system. 
Our system considers the transmission of multiple codewords from two binary block codes $\mathcal{C}_{1}$ and $\mathcal{C}_{2}$, each of length $N$.
In this study, we use two different LDPC codes from the same ensemble.
At the transmitter, encoded codeword matrices $\bm{C}_{1}$ and $\bm{C}_{2}$ are processed through shifting, bit-to-base mapping, and interleaving to form the transmitted matrix $\bm{X}$.
At the receiver, codewords in $\bm{C}_{1}$ and $\bm{C}_{2}$ are estimated sequentially, involving detection for recovering ID errors and cooperative decoding for correcting substitution errors and sequence losses.
Estimated codewords in each round inform subsequent rounds.
\subsection{Encoding Scheme}
We assume that the code length is given by $N=msd$, where $m$, $s$, and $d$ are positive integers. 
For each $u \in \{1,2\}$, the codeword matrix $\bm{C}_{u} \in \Sigma_{2}^{ms \times Ld}$ is denoted as
\begin{align}
  \bm{C}_{u}=
    \begin{bmatrix}
      \bm{C}_{u}^{(1,1)} & \bm{C}_{u}^{(1,2)} & \cdots & \bm{C}_{u}^{(1,L)}\\
      \bm{C}_{u}^{(2,1)} &  \bm{C}_{u}^{(2,2)} & \cdots       & \bm{C}_{u}^{(2,L)}\\
      \vdots & \vdots         & \ddots & \vdots \\
      \bm{C}_{u}^{(m,1)} & \bm{C}_{u}^{(m,2)} & \cdots & \bm{C}_{u}^{(m,L)}
    \end{bmatrix},
\end{align}
where each submatrix $\bm{C}_{u}^{(i,j)}$ is of size $s \times d$.
The $j$-th codeword corresponds to the block column matrix $\bm{C}_{u}^{(j)}$ of size $ms \times d$;
that is, $L$ codewords are used to form $\bm{C}_{u}$.
The rest of encoding process consists of three stages: shifting, mapping, and interleaving:
\begin{figure}[t]
  \centering
  \includegraphics[width=.78\linewidth]{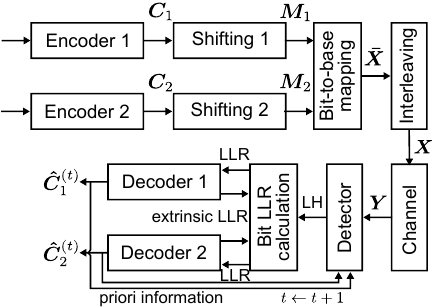}
  \caption{Diagram of the proposed system based on the SI coding scheme.}
  \label{fig:system}
\end{figure}
\subsubsection{Shifting}
For each $u \in \{1, 2\}$, with the shift vector $\bm{T}_{u} = (T_{u,1}, T_{u,2}, \ldots, T_{u,m}) \in [0:T_{\rm max}]^{m}$, each submatrix $\bm{C}_{u}^{(i,j)}$ is shifted to either a higher or the same column number to form the intermediate binary matrix $\bm{M}_{u} \in \Sigma_{2}^{ms \times L'd}$, where $T_{\rm max}$ is the maximum shift value and $L'$ is given by $L + T_{\rm max}$.
Specifically, for each pair $(i,j) \in [1:m] \times [1:L]$, the submatrix $\bm{M}_{u}^{(i,j)}$ is given by
\begin{align}
  \bm{M}_{u}^{(i,j)} = \bm{C}_{u}^{(i,j-T_{u,i})},
\end{align}
if $1 \leq j-T_{u,i} \leq L$.
Otherwise, $\bm{M}_{u}^{(i,j)}$ becomes a matrix padded with i.i.d.\ equiprobable bits.
These padding bits are shared between the transmitter and receiver.
\begin{exam}\label{ex1}
  With $\bm{T}_{u}= (0,1,2,3)$ and $L=3$, we have
  \begin{align*}
    \bm{M}_{u} =
    \begin{bmatrix}
      \bm{C}_{u}^{(1,1)} & \bm{C}_{u}^{(1,2)} & \bm{C}_{u}^{(1,3)} & \ast               & \ast               &\ast                \\
      \ast               & \bm{C}_{u}^{(2,1)} & \bm{C}_{u}^{(2,2)} & \bm{C}_{u}^{(2,3)} & \ast               &\ast                \\
      \ast               & \ast               & \bm{C}_{u}^{(3,1)} & \bm{C}_{u}^{(3,2)} & \bm{C}_{u}^{(3,3)} &\ast               \\
      \ast               & \ast               & \ast               & \bm{C}_{u}^{(4,1)} & \bm{C}_{u}^{(4,2)} &\bm{C}_{u}^{(4,3)} 
    \end{bmatrix},
  \end{align*}
  where $\ast$ stands for the padding bits.
\end{exam}
\subsubsection{Mapping}
Next, the intermediate quaternary matrix $\bm{\bar{X}} \in \Sigma_{4}^{ms \times L'd}$ is obtained by mapping from $\bm{M}_{1}$ and $\bm{M}_{2}$.
As the bit-to-base mapping $\psi$, we employ \textit{quaternion coding}, where a two-bit pair is assigned to a DNA base. 
The mapping $\psi: \Sigma_{2}^{2} \rightarrow \Sigma_{4}$ is defined by the relations $(0,0) \mapsto A$, $(0,1) \mapsto C$, $(1,0) \mapsto T$, and $(1,1) \mapsto G$.
Using $\bm{M}_{u}$ and $\psi$, each element $\bar{x}(k,z)$ of $\bm{\bar{X}}$ is given by
\begin{align}
  \bar{x}(k,z) = \psi \left( m_{1}(k,z) , m_{2}(k,z) \right).
\end{align}
For later use, we define $\psi^{-1}_{u}(v) = b_{u}$ for each $u \in \{1,2\}$ given that $\psi(b_{1}, b_{2}) = v$.
\subsubsection{Interleaving}
Finally, the transmitted matrix $\bm{X} \in \Sigma_{4}^{ms \times L'd}$ is generated by performing a row permutation on $\bm{\bar{X}}$.
Specifically, $\bm{X}$ is computed as $\bm{X}=\bm{P}\bm{\bar{X}}$, where $\bm{P}=(\bm{P}^{(1)},\bm{P}^{(2)},\ldots, \bm{P}^{(m)})^{\top}$ is a permutation matrix of size $ms \times ms$.
For $i \in [1:m]$, each element $p^{(i)}(k,z)$ of the $s \times ms$ submatrix $\bm{P}^{(i)}$ is given by
\begin{align}
  p^{(i)}(k,z) =
  \begin{cases}
    1 & \text{if } z = (k-1)m+i, \\
    0 & \text{otherwise.}
  \end{cases}
\end{align}
\begin{exam}\label{ex2}
  With $\bm{T}_{1}= \bm{T}_{2} = (0,1,2,3)$, $s=2$, $d=1$, and $L=3$, we have
  \begin{align*}
    \bm{X} =
    \begin{bmatrix}
      \bar{x}(1,1) & \bar{x}(1,2) & \bar{x}(1,3) & \ast         & \ast         &\ast         \\
      \ast         & \bar{x}(3,2) & \bar{x}(3,3) & \bar{x}(3,4) & \ast         &\ast         \\
      \ast         & \ast         & \bar{x}(5,3) & \bar{x}(5,4) & \bar{x}(5,5) &\ast         \\
      \ast         & \ast         & \ast         & \bar{x}(7,4) & \bar{x}(7,5) &\bar{x}(7,6) \\
      \bar{x}(2,1) & \bar{x}(2,2) & \bar{x}(2,3) & \ast         & \ast         &\ast         \\
      \ast         & \bar{x}(4,2) & \bar{x}(4,3) & \bar{x}(4,4) & \ast         &\ast         \\
      \ast         & \ast         & \bar{x}(6,3) & \bar{x}(6,4) & \bar{x}(6,5) &\ast         \\
      \ast         & \ast         & \ast         & \bar{x}(8,4) & \bar{x}(8,5) &\bar{x}(8,6)
    \end{bmatrix},
  \end{align*}
  where $\ast$ stands for symbols from the padding bits.
  Each column vector $\bm{x}_{j}$ corresponds to a DNA sequence.
\end{exam}
In the SI coding scheme, each codeword is dispersed across multiple DNA sequences, thereby limiting the impact of sequence loss to only parts of a codeword.
Furthermore, the known (padding) symbols within DNA sequences (e.g., the $4$ and $8$-th symbols in Ex.~\ref{ex2}) serve as synchronization markers\cite{marker}, aiding in the correction of ID errors.
\subsection{Decoding Scheme}
\begin{figure}[!t]
  \begin{algorithm}[H]
  \caption{Decoding algorithm}
  \label{alg:alg_dec}
  \begin{algorithmic}[1]
    \REQUIRE Channel output matrix $\bm{Y}$
    \ENSURE Estimated codeword matrices $\bm{\hat{C}}_{1}$ and $\bm{\hat{C}}_{2}$
    \STATE Initialize $\bm{Q}_{u}$ and $\bm{L}_{u}$ for $u \in \{1,2\}$, and also initialize $\bm{\Gamma}$ and $\bm{\Pi}$
    \FOR {$t = 1$ to $L$}
      \FOR {$i =0$ to $T_{\rm max}$}
        \FOR {$j \in \mathcal{I}_{t+i}$}
          \IF{$\bm{y}_{j} \neq \mathord{?}$}
          \STATE For $\bm{y}_{j}$, execute detection to compute $\bm{\gamma}_{j}$
          \ENDIF
        \ENDFOR
      \ENDFOR
      \STATE $\ell = 0$
      \WHILE{($\bm{\hat{C}}_{1}^{(t)} \notin \mathcal{C}_{1}$ or $\bm{\hat{C}}_{2}^{(t)} \notin \mathcal{C}_{2}$) and $\ell < \ell_{\rm max}$}
        \FOR {$u \in \{1,2\}$}
          \STATE Compute $\bm{L}_{u}^{(t)}$ based on $\{\bm{\Gamma}^{(i,t+T_{u,i})}\}_{i \in [1:m]}$ and $\bm{Q}_{\bar{u}}^{(t)}$, where $\bar{u}\neq u$.
          \STATE Execute a single iteration of SP decoding to update $\bm{\hat{C}}_{u}^{(t)}$ and $\bm{Q}_{u}^{(t)}$
        \ENDFOR  
        \STATE $\ell \gets \ell + 1$
      \ENDWHILE
      \FOR {$u \in \{1,2\}$}
        \STATE Update $\{\bm{\Pi}^{(i,t+T_{u,i})}\}_{i \in [1:m]}$ based on $\bm{Q}_{1}^{(t)}$ and $\bm{Q}_{2}^{(t)}$
      \ENDFOR 
    \ENDFOR
  \end{algorithmic}
  \end{algorithm}
\end{figure}
For simplicity, we define the set of column indices corresponding to $\bm{X}^{(t)}$ (and $\bm{\bar{X}}^{(t)}$) of size $ms \times d$ as $\mathcal{I}_{t}$ for $t \in [1:L']$ and assume that $\{j \in [1:m] \mid T_{u,j} = i\} \neq \emptyset$ for all $i\in [0:T_{\rm max}]$ and $u\in \{1,2\}$.
For $u \in \{1,2\}$, $\bar{u}$ represents the complement of $u$.
Let $\bm{L}_{u}$ and $\bm{Q}_{u}$ represent the decoder \textit{input (a priori)} and \textit{output (extrinsic)} log-likelihood ratio (LLR) matrices, respectively, each of size $ms \times Ld$, associated with $\bm{C}_{u}$.
Moreover, $\bm{\Pi}$ and $\bm{\Gamma}$ represent the detector \textit{input} and \textit{output} probability matrices, respectively, each of size $ms \times L'd$ and related to $\bm{\bar{X}}$.
In these matrices, every element, $\pi(k,z)$ for $\bm{\Pi}$ and $\gamma(k,z)$ for $\bm{\Gamma}$, is a quaternary tuple.
The element of each tuple, $\pi(k,z,v)$ and $\gamma(k,z,v)$, represents respectively as $p(\bar{x}(k,z)=v)$ and $p(\bm{Y} \mid \bar{x}(k,z)=v)$, where $v\in \Sigma_{4}$.
The $s \times d$ submatrix representations, $\bm{L}_{u}^{(i,j)}$, $\bm{Q}_{u}^{(i,j)}$, $\bm{\Pi}^{(i,j)}$, and $\bm{\Gamma}^{(i,j)}$, are used for $\bm{L}_{u}$, $\bm{Q}_{u}$, $\bm{\Pi}$, and $\bm{\Gamma}$, respectively.

Algorithm~\ref{alg:alg_dec} provides an entire procedure to decode the codeword matrices $\bm{C}_{1}$ and $\bm{C}_{2}$.
Focusing on the $t$-th estimation round to obtain the estimates $\bm{\hat{C}}_{1}^{(t)}$ and $\bm{\hat{C}}_{2}^{(t)}$, the proposed scheme consists of two main stages: detection and decoding.
\subsubsection{Detection (lines 3--9 in Alg.\ \ref{alg:alg_dec})}
The ID errors in each received sequence $\bm{y}_{j}$ are inferred by the detector whenever $\bm{y}_{j} \neq \mathord{?}$.
More precisely, the detector computes the corresponding LHs $\bm{\gamma}_{j}$.
The sequence $\bm{y}_{j}$ can be modeled as the output of a hidden Markov model (HMM)\cite{marker,water,my_irr}.
Similar to \cite{my_tcom23,my_irr}, we adopt a channel state process characterized by a random walk with reflecting boundaries over the set $[-D_{\rm max}:+D_{\rm max}]$, where $D_{\rm max}$ represents the maximum drift between IDS channel input and output sequences.
The detection task is accomplished by applying the forward-backward (FB) algorithm to a HMM channel graph.

In our scheme, the $t$-th codewords $\bm{C}_{u}^{(t)}$ are spread across $\{\bm{X}^{(t+i)}\}_{i \in [0:T_{\rm max}]}$.
Consequently, in order to obtain the corresponding LH submatrices $\{\bm{P}\bm{\Gamma}^{(t+i)}\}_{i \in [0:T_{\rm max}]}$, the FB algorithm is performed on each received column vector $\bm{y}_{j}$ for all $j \in \mathcal{I}_{t+i}, i \in [0:T_{\rm max}]$, whenever $\bm{y}_{j} \neq \mathord{?}$.
When $\bm{y}_{j} = \mathord{?}$, $\bm{x}_{j}$ is treated as erased symbols.
For $u \in \{1,2\}$ and $i \in [1:m]$, $\bm{\Gamma}^{(i, t + T_{u,i})}$ is associated with $\bm{C}_{u}^{(i,t)}$.

\begin{figure}[t]
  \centering
  \includegraphics[width=.9\linewidth]{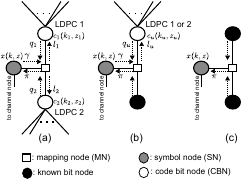}
  \caption{Illustration of message passing around a mapping node.}
  \label{fig:nodeupd}
\end{figure}

Using a factor graph representation, as shown in Fig.\ \ref{fig:nodeupd}, each transmitted symbol node $x(k,z)$ is connected to a node on the channel graph and a mapping node (MN).
The MN, besides $x(k,z)$, is connected to two other nodes, which are either code bit nodes (CBNs) or known bit nodes (KBNs).
The detector input priori probability $\pi(k,z)$ in $\bm{\Pi}$ is provided from the MN. 
Since known and pre-decoded bit nodes can provide reliable information, they function as synchronization markers in the detection process.
\subsubsection{Decoding (lines 11--17 in Alg.\ \ref{alg:alg_dec})}
Each LDPC decoder operates based on the standard log-domain sum-product (SP) decoding algorithm, utilizing a so-called flooding schedule.
As shown in Fig.\ \ref{fig:nodeupd}, each CBN is connected to a MN, and they exchange messages with each other.
This means that the two decoders can operate cooperatively.
A single iteration in the decoding process comprises the following steps:
\begin{enumerate}
  \item Update LLRs $l_{u}(k_{u},z_{u})$ in $\bm{L}_{u}^{(t)}$ for $u \in \{1,2\}$.
  \item Update CBN-to-Check Node (CN) messages in each LDPC Tanner graph.
  \item Update CN-to-CBN messages in each LDPC Tanner graph.
  \item Update estimates $c_{u}(k_{u},z_{u})$ in $\bm{\hat{C}}_{u}^{(t)}$ and $q_{u}(k_{u},z_{u})$ in $\bm{Q}_{u}^{(t)}$ for $u \in \{1,2\}$.
\end{enumerate}
The calculation of each LLR value $l_{u}(k_{u},z_{u})$ in Step 1) is classified into three cases, depending on a shift vector $\bm{T}_{u}$.
The first case occurs at positions $i$ where $T_{1,i} = T_{2,i}$, indicating that both CBNs $c_{1}(k_{1},z_{1})$ and $c_{2}(k_{2},z_{2})$ are bits of the $t$-th codewords.
In this case, the corresponding LLR value $l_{u}(k_{u},z_{u})$ for $u \in \{1,2\}$ is computed in each iteration as follows
\begin{align}\label{eq:ltollr1}
  l_{u}(k_{u},z_{u}) = \ln \frac{\displaystyle\sum_{\psi^{-1}_{u}(v)=0, v \in \Sigma_{4}} \gamma(k,z,v) f(\psi^{-1}_{\bar{u}}(v), q_{\bar{u}}(k_{\bar{u}},z_{\bar{u}})) }{\displaystyle\sum_{\psi^{-1}_{u}(v)=1, v \in \Sigma_{4}} \gamma(k,z,v)f(\psi^{-1}_{\bar{u}}(v), q_{\bar{u}}(k_{\bar{u}},z_{\bar{u}})) }.
\end{align}
The function $f(b, q)$ is defined as
\begin{align}
  f(b, q) =
  \begin{cases}
    \frac{\exp(q)}{1+\exp(q)} & \text{if } b = 0, \\
    \frac{1}{1+\exp(q)} & \text{otherwise.}
  \end{cases}
\end{align}
The second case occurs at positions $i$ where $T_{1,i} \neq T_{2,i}$, and a CBN $c_{u}(k_{u},z_{u})$ is a bit of the $t$-th codeword while CBN $c_{\bar{u}}(k_{\bar{u}},z_{\bar{u}})$ ($\bar{u} \neq u$) is not.
In this case, $l_{u}(k_{u},z_{u})$ is calculated using Eq.\ (\ref{eq:ltollr1}).
If $c_{\bar{u}}(k_{\bar{u}},z_{\bar{u}})$ is a pre-decoded CBN (i.e., $T_{u,i} < T_{\bar{u},i}$), the LLR computation is enhanced. 
On the other hand, if it is a non-decoded CBN (i.e., $T_{u,i} > T_{\bar{u},i}$), unreliable LLR computation is performed.
The third case also occurs at positions $i$ where $T_{1,i} \neq T_{2,i}$, and the MN is connected to a CBN and a KBN.
In this case, $l_{u}(k_{u},z_{u})$ is given by
\begin{align}\label{eq:ltollr2}
  l_{u}(k_{u},z_{u}) = \ln \frac{\displaystyle\sum_{\psi^{-1}_{u}(v)=0,\psi^{-1}_{\bar{u}}(v)=s,v \in \Sigma_{4}} \gamma(k,z,v)}{\displaystyle\sum_{\psi^{-1}_{u}(v)=1,\psi^{-1}_{\bar{u}}(v)=s,v \in \Sigma_{4}} \gamma(k,z,v)},
\end{align}
where $s$ represents the known bit value.
For cases 2 and 3, it is sufficient to compute the LLR values only in the first iteration.

The above procedure continues until valid codewords are obtained or a maximum number of iterations is reached.
Afterwards, the priori probability $\pi(k,z)$ is updated using the decoder output extrinsic LLR $q_{u} (k_{u},z_{u})$ in $\bm{Q}_{u}^{(t)}$ (see lines 18--20 in Alg.\ \ref{alg:alg_dec}), which helps the subsequent estimation rounds.
When the MN is connected to two CBNs, $\pi(k,z,v)$ is calculated as follows
\begin{align}
  \pi(k,z,v) = \prod_{u \in \{1,2\}} f(\psi^{-1}_{u}(v), q_{u}(k_{u},z_{u})) \label{eq:llrtol1}
\end{align}
When the MN is connected to a CBN and a KBN, the calculation is as follows
\begin{align} 
  \pi(k,z,v) = 
  \begin{cases}
    f(\psi^{-1}_{u}(v), q_{u}(k_{u},z_{u})) & \text{if } \psi^{-1}_{\bar{u}}(v)=s, \\
    0 & \text{otherwise.}
  \end{cases}
\end{align}
where $s$ represents the known bit value.
\begin{figure*}[t]
  \begin{minipage}[b]{0.33\hsize}
  \centering
  \includegraphics[width=.99\linewidth]{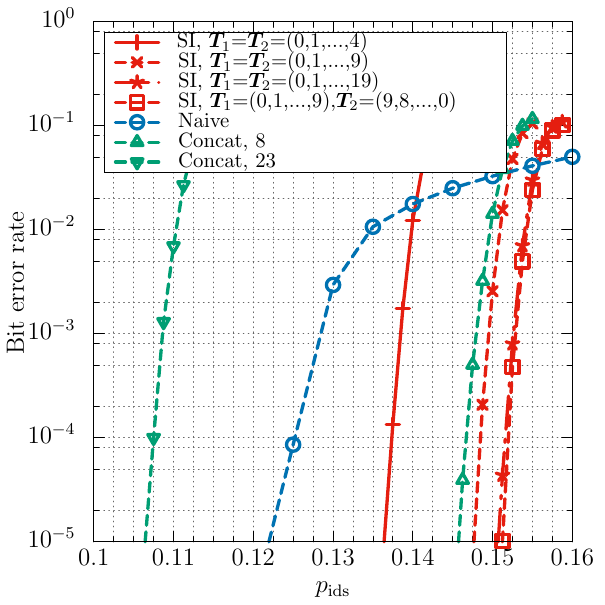}
  \subcaption{$p_{\rm e}=0$}\label{res11}
  \end{minipage}
  \begin{minipage}[b]{0.33\hsize}
  \centering
  \includegraphics[width=.99\linewidth]{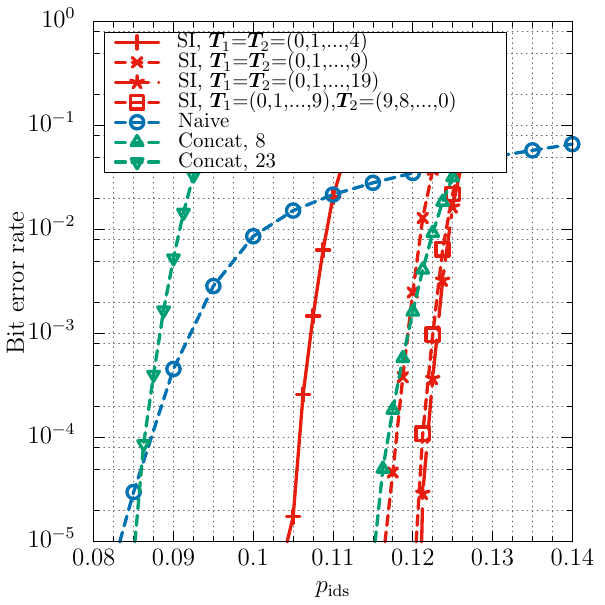}
  \subcaption{$p_{\rm e}=0.1$}\label{res12}
  \end{minipage}
  \begin{minipage}[b]{0.33\hsize}
    \centering
    \includegraphics[width=.99\linewidth]{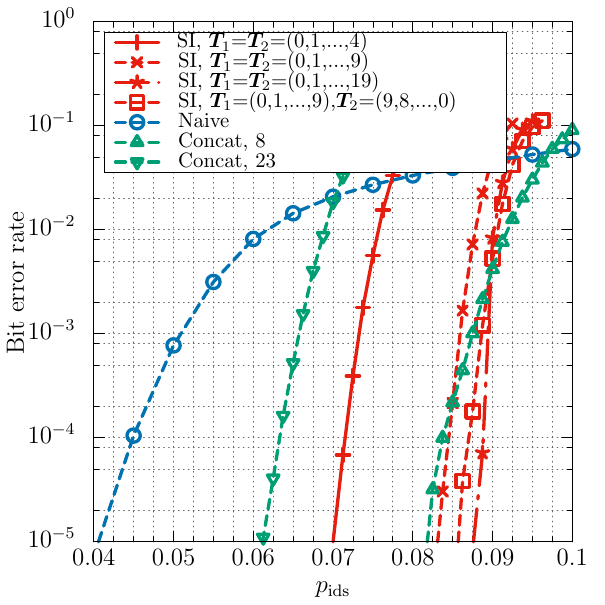}
    \subcaption{$p_{\rm e}=0.2$}\label{res13}
  \end{minipage}
  \caption{BER curves of the SI coding scheme, naive separation scheme, and concatenation scheme, where $n=100$ and $d=1,000$.}
  \label{fig:res1}
\end{figure*}
\subsection{Efficiency and Complexity}
We first show the efficiency of the SI coding scheme.
Due to the allocation of known-bits in the transmitted matrix $\bm{X}$, there is an associated rate loss.
Given a code rate $R_{u}$ of a component code $\mathcal{C}_{u}$ for $u\in \{1,2\}$ and the number of codewords $L$,
the transmission rate $R_{\rm tx}$ (bits per base) is given by
\begin{align}
  R_{\rm tx} = \left(\frac{L}{L+T_{\rm max}} \right) \left( R_{1}+R_{2} \right).
\end{align}
Obviously, $R_{\rm tx}$ approaches $R_{1}+R_{2}$ for a large $L$ ($L \gg T_{\rm max}$).

We next address the computational complexity.
Since the detection is dominant in the overall computational process, we focus on this aspect here.
To decode $\bm{C}_{1}^{(t)}$ and $\bm{C}_{2}^{(t)}$ in each estimation round, the FB algorithm is applied to at most $\sum_{i=0}^{T_{\rm max}}|\mathcal{I}_{t+i}| = (T_{\rm max}+1)d$ sequences before the decoding stage.
The complexity of the FB algorithm for an IDS channel is proportional to $D \triangleq 2D_{\rm max}+1$, multiplied by the number of timesteps\cite{water}.
Consequently, the complexity of detection per round is limited to $O((T_{\rm max}+1)dnD) = O((T_{\rm max}+1)ND)$.
Note that each FB algorithm instance can be performed in parallel.
\section{Numerical Results}\label{sec:sim}
We present the finite-length performance of our proposed scheme.
We focus on the $t$-th estimation round, under the assumption that all preceding $t-1$ stages have been successfully completed.
Throughout this section, for a given total IDS error probability $p_{\rm ids}$, the insertion, deletion, and substitution probabilities are set at $p_{\rm i}=0.17p_{\rm ids}$, $p_{\rm d}=0.40p_{\rm ids}$, and $p_{\rm s}=0.43p_{\rm ids}$, respectively, according to the statistical results in \cite{HEDGES}.
Moreover, we choose the maximum drift to accomodate five times larger than the standard deviation of the average drift size at position $n$,
i.e., $D_{\rm max} = |n(p_{\rm i}-p_{\rm d})| + 5\sqrt{n(p_{\rm i} + p_{\rm d}-(p_{\rm i}-p_{\rm d})^{2})}$.

We employ various types of shift vectors, primarily with $T_{\rm max}=9$, in conjunction with two LDPC codes, each of length $N = 100,000$, drawn from a rate-$R_{u} = 0.5$ LDPC code ensemble specifically designed for additive white Gaussian noise (AWGN) channels \cite{ldpcopt}.
We compare our scheme against the following baselines:
\begin{enumerate}
    \item A naive codeword separation scheme, equivalent to our scheme with $\bm{T}_{1}=\bm{T}_{2}=(0)$. We employ a rate-$0.5$ LDPC code ensemble designed for IDS channels \cite{my_irr}.
    \item A codeword separation scheme with concatenated LDPC and marker codes \cite{dna_ldpcmk1, dna_ldpcmk2}. This scheme utilizes the rate-$0.5$ AWGN-LDPC code ensemble.
    For markers, two random symbols from $\Sigma_{4}$ are inserted every $8$ or $23$ symbols into each DNA base sequence formed from an LDPC codeword.
    As a result, the overall coding rates are $0.40$ and $0.46$, respectively.
\end{enumerate}
Both schemes use iterative detection and decoding, where the first $10$ iterations include detection at each SP decoder iteration.
This corresponds to the detection complexity of the proposed scheme with $T_{\rm max}=9$.
The maximum number of iterations, $\ell_{\rm max}$, is set to $100$.

Figure~\ref{fig:res1} illustrates the bit error rate (BER) curves as a function of $p_{\rm ids}$ for a DNA channel with $p_{\rm e} \in \{0,0.1,0.2\}$, where $n=100$ and $d=1,000$.
From the figure, it can be seen that the SI coding scheme achieves impressive performance, characterized by a sharp waterfall drop in BER.
Specifically, the SI coding scheme outperforms other schemes across all values of the block erasure probability $p_{\rm e}$, demonstrating its effectiveness against both IDS errors and sequence losses. 
Moreover, an increase in $T_{\rm max}$ tends to enhance decoding performance, albeit at the expense of increased detection cost.
Additionally, performance is influenced by the choice of shift vectors $(\bm{T}_{1}, \bm{T}_{2})$, even when $T_{\rm max}$ is fixed at $9$.
\section{Conclusion}\label{sec:conc}
In this study, we proposed the SI coding scheme for correcting IDS errors and sequence losses in DNA-based data storage systems.
The proposed scheme utilizes a three-stage encoding process at the transmitter, where multiple codewords from two LDPC codes are appropriately rearranged to generate DNA base sequences embedded with a marker structure.
At the receiver, an efficient decoding strategy is employed to sequentially estimate the original codewords, leveraging previously decoded codewords to assist in the detection process.
Simulation results demonstrated that the proposed scheme achieves excellent performance over IDS errors and sequence losses.
In future work, we plan to conduct achievable information rate analysis and design a pair of shift vectors $(\bm{T}_{1},\bm{T}_{2})$ to improve decoding performance while reducing complexity.
\section*{Acknowledgments}
This work was supported by JSPS KAKENHI Grant Number JP21K14160.

\end{document}